# Joint Radar-Communication Waveform Design Based on Composite Modulation

Yuan Quan, Fulai Wang, Longfei Shi and Jiazhi Ma

*Abstract*—Joint radar-communication (JRC) waveform can be used for simultaneous radar detection and communication in the same frequency band. However, radar detection processing requires the prior knowledge of the waveform including the embedded information for matched filtering. To remove this requirement, we propose a unimodular JRC waveform based on composite modulation where the internal modulation embeds information by mapping the bit sequence to different orthogonal signals, and the external modulation performs phase modulation on the internal waveform to satisfy the demand of detection. By adjusting the number of the orthogonal signals, a trade-off between the detection and the communication performance can be made. Besides, a new parameter, dissimilarity, is defined to evaluate the detection performance robustness to unknown embedded information. The numerical results show that the SER performance of the proposed system is similar to that of the multilevel frequency shift keying system, the ambiguity function resembles that of the phase coded signal, and the dissimilarity performance is better than other JRC systems.

## I. INTRODUCTION

BOTH high-resolution detection and high-speed communication have an increasing demand for spectrum resources, which exacerbates the congestion of the electromagnetic spectrum. Hence, there are numerous researchers focusing on the sharing of the spectrum between radar and communication systems.

Generally, in an overlapping frequency band, radar and communication will be the interference of the other. To solve this problem, traditional methods divide resources such as spectrum, space, time and polarization into different systems [1]. However, such division is inefficient because of the lack of collaboration among systems. Further, coexistence and cooperation methods improve the performance by dynamic spectrum allocation and interference mitigation [2][3]. Nevertheless, additional resources will be consumed and the degree of freedom will be limited. Instead, codesign methods realize both detection and communication in one system and investigate the probability to maximize the spectrum efficiency [4]. Codesign methods include but are not limited to joint radar-communication (JRC) systems based on multi-input-multi-output system, beamforming technology and JRC waveform design [5]-[7]. This paper will focus on the JRC waveform design which aims to embed information symbols into the radar waveform or detect targets with the communication waveform.

A number of waveforms have been investigated and applied to JRC systems. Orthogonal frequency division multiplex (OFDM) waveform has been widely used in high-speed communication and high-resolution sensing [8]. However, the high peak to average power ratio (PAPR) is a problem for radar application [9]. Constant envelope OFDM signal is designed to solve this problem, but the peak sidelobe level (PSL) and the range resolution will be sacrificed [10]. Besides, Liner frequency modulation (LFM) signal which is widely used in the radar system and the spread spectrum communication system, has been investigated by many researchers [11]. In [12], the information sequence is embedded into the LFM signal by reduced binary phase shift keying (BPSK), and the trade-off between the ambiguity function and the bit error rate (BER) performance is made by the adjustment of the degree of the reduction. To improve the spectrum efficiency, information sequence is modulated with continuous phase modulation and phase-attached to a polyphase-coded frequency-modulated (PCFM)-LFM radar waveform [13][14], and the modulation index is used to balance the radar and the communication performance in [15]. To increase the amount of information embedded in the waveform, the approach investigated in [16] embeds information symbol by combining the communication frequency modulation term (CFMT) with radar LFM and balance the performance with the weighted coefficient.

JRC waveforms mentioned above can be applied in monostatic broadcast channel topology where the transmitted waveform including the embedded information, is fully known to the radar receiver [3]. However, the detection performance of them degrades in bistatic broadcast channel topology where the embedded information is unknown to the radar receiver, and there will be a mismatch during the matched filtering [3]. Demonstrated in Figure 1, the JRC waveform will be transmitted through the reflect line if the direct line between the two users is obstructed by geographical factors such as buildings, hills and earth curvature, which exacerbates the multi-path effect and raises the symbol error rate (SER) [17].

This work was supported in part by the National Natural Science Foundation of China (NSFC) under Grant 61871385.

The authors are with the State Key Laboratory of Complex Electromagnetic Environment Effects on Electronics and Information System, National University of Defense Technology, Changsha, 410073, China. (e-mail: qy0690@gmail.com; wflmadman@outlook.com; longfeishi@sina.com; jzmanudt@163.com).

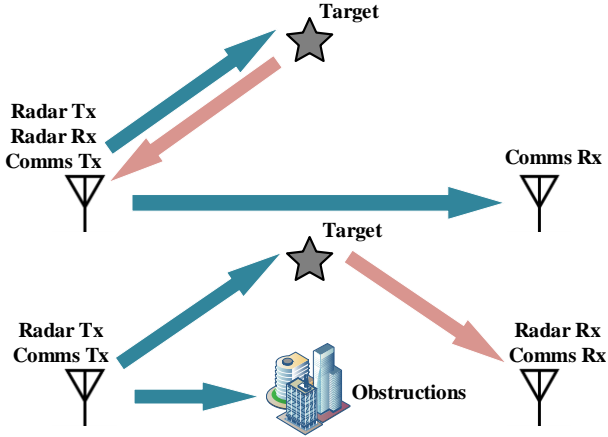

Fig. 1. Comparison between monostatic broadcast channel topology (above) where one of the users acts as a monostatic radar and a communication transmitter and the other acts as a communication receiver and bistatic broadcast channel topology (below) where one of the users acts as a radar transmitter and a communication transmitter and the other acts as a radar receiver and a communication receiver.

To overcome the multi-path effect, the channel state information (CSI) which is acquired by the channel estimation is essential. It is noteworthy that the CSI is required for both detection and communication purposes. Therefore, the redundancy can be utilized to design an integrated detection process, which helps to maximize the spectrum efficiency [18], and this process should be finished before the information demodulation. Correspondingly, the channel detection performance must show robustness to unknown information (RUI), which represents that the information embedded in the waveform is not required for the detection process, and the detection performance is robust to the unknown embedded information.

Possible solutions for the RUI can be found in communication-based JRC waveforms. In [18], the training signal is modified for target detection, and the detection performance and the communication performance are balanced by the power allocation. Nevertheless, the inconstant power leads to an inconstant waveform envelope which is a problem for radar system. Instead of allocating the power, the trade-off is made by changing the length of the training signal in [19], and the problem of inconstant envelope is solved. Both of these methods divide the time domain into the training signal part and the data signal part. Therefore, the detection performance is robust to the unknown transmitted information. However, guard interval (zero sequence) is embedded between the training signal and the data signal for the sake of decoupling, and the power of the transmitter will be wasted during the interval. Besides, the channel characteristic changes between the training signal and the data signal, and this difference may reduce the channel equalization performance.

In conclusion, JRC waveforms mentioned above are available in many scenarios but limited in the bistatic broadcast channel topology in Figure 1. Thus, a novel JRC waveform which satisfies the following requirements is proposed:
- Simultaneous detection and communication in the same band.
- Constant envelope of the transmitted waveform.
- Detection performance robustness to unknown information.
- Trade-off between radar and communication performance.

Composite modulation is designed for the proposed waveform. For the internal modulation, a group of orthogonal signals are used to map to different information sequences. For the external modulation, the external phase (EP) is modulated on the internal waveform to optimize the PSL and the Doppler performance. The envelope of the proposed waveform will be constant if that of the orthogonal signals is constant since the external modulation only changes the phase of the internal waveform. The RUI is realized by the design of the corresponding radar signal processing which does not require the knowledge of the embedded information. The trade-off between the amount of embedded information and the detection performance is made by changing the number of the orthogonal signals. The performance of the proposed waveform, including the detection performance, the Doppler performance, the symbol error rate (SER) and the RUI performance are analyzed and simulated.

## II. WAVEFORM DESIGN BASED ON COMPOSITE MODULATION

### A. Properties of orthogonal signals

Define $\{s_k(t)\}_{k=1}^{K}$ as $K$ orthogonal signals with the signal energy of $E_s$. The cross-correlation of $s_i(t)$ and $s_j(t)$ can be expressed as

$$R_{ij}(t) = s_i(t) \otimes s_j^*(-t), i,j = 1,...,K \quad (1)$$

where $\otimes$ and $(\cdot)^*$ represents the convolution operation and the conjugation operation, respectively. When $i=j$, (1) becomes the autocorrelation of $s_i(t)$. According to [20], if the signals are ideal orthogonal, the autocorrelation and the cross-correlation can be described as

$$R_{ii}(t) = \begin{cases} E_s, t = 0 \\ 0, t \neq 0 \end{cases} \quad (2)$$

and

$$R_{ij}(t) = R_{ji}^*(-t) = 0, \forall t, i \neq j. \quad (3)$$

To simplify the following discussion, we will deduce the process in the situation of ideal orthogonality in this section. However, the ideal orthogonality cannot be realized but approximated in real systems. In Section III, the influence of the nonideal orthogonality will be discussed.

### B. Transmitted waveform model

The process of waveform generation is shown in Figure 2. Specifically, the transmitted waveform can be expressed as

$$s(t) = \sum_{k=1}^{K}\sum_{n=0}^{N-1} c_{n,k} s_k(t - nT_s) e^{j\varphi_n} \quad (4)$$

where

$$c_{n,k} = \begin{cases} 1, I_n = [(k-1)_{10}]_2 \\ 0, I_n \neq [(k-1)_{10}]_2 \end{cases}, \quad (5)$$

in which $[(\cdot)_{10}]_2$ denotes the conversion from decimal numbers to binary numbers, $I_n = \overline{I_{n,1} \ldots I_{n,l} \ldots I_{n,L}}$ represents the binary number embedded in the $n$-th information symbol, $I_{n,l}$ is the $l$-th bit ('0' or '1') of the $n$-th binary number, $L=\log_2 K$ represents the number of bits in one information symbol, $N$ is the number of the information symbols in the waveform, $T_s$ is the duration of the orthogonal signals, and $\varphi_n$ is the EP of the $n$-th information symbol. The duration of the waveform is $T=NT_s$ and the energy of $s(t)$ is $E=NE_s$. The mapping relationship is demonstrated in Table 1. This mapping process is defined as the internal modulation. The internal waveform consists of $N$ orthogonal signals which are chosen from $s_1$ to $s_K$, and each signal will be multiplied by an EP. This multiplication process is defined as the external modulation.

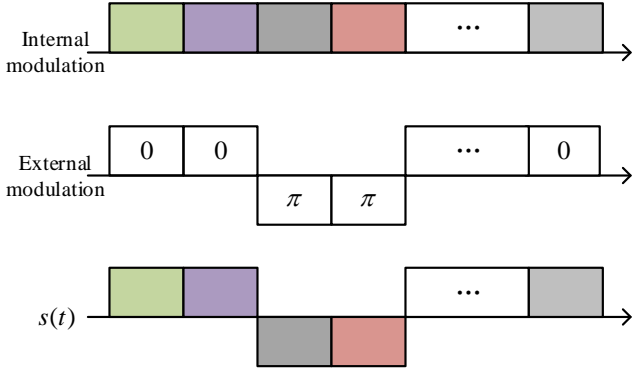

Fig. 2. Transmitted waveform model. Each color represents an orthogonal signal from the $K$ orthogonal signals.

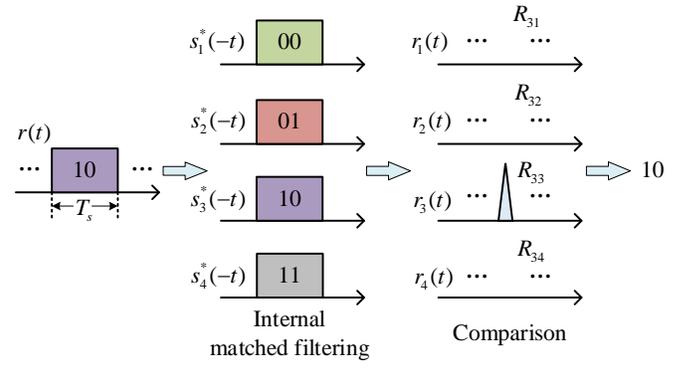

Fig. 3. An example of the information demodulation process. The number of the orthogonal signals is $K=4$ and the embedded binary number is $I_n$=10. Only one information symbol is shown in the figure. The cross-correlations are ignored according to (3).

Table. 1. The mapping relationship between binary sequences and orthogonal signals.

|  | Signal 1 | Signal 2 | … | Signal $K$ |
|---|---|---|---|---|
| $I_n$ | 00…00 | 00…01 | … | 11…11 |

### C. Information demodulation and radar signal processing

Considering a single point stationary target, the received signal can be expressed as

$$r(t) = s(t-t_0) + w(t) \quad (6)$$

where $w(t)$ is the additive white Gaussian noise whose power spectral density is $N_0/2$ and $t_0$ is the time delay. Then, the internal matched filtering is performed by matching the received signal with $K$ matched filters. The output of the $k$-th matched filter can be written as

$$r_k(t) = r(t) \otimes s_k^*(-t), k=1,...,K. \quad (7)$$

Substituting (1) and (4) into (7), it can be obtained that

$$r_k(t) = \sum_{n=0}^{N-1}\sum_{j=1}^{K} c_{n,j} R_{jk}(t-t_0-nT_s) e^{j\varphi_n} \\ + w_k(t), k=1,...,K \quad (8)$$

where

$$w_k(t) = w(t) \otimes s_k^*(-t). \quad (9)$$

It has been illustrated in Section II-B that only one orthogonal signal is transmitted during a symbol duration $T_s$. Accordingly, one of the outputs of the $K$ matched filters will be the autocorrelation of the transmitted signal, and others will be the cross-correlations of the transmitted signal and other signals. Displayed in Figure 3, by finding the maximum of the $K$ outputs, the serial number of the transmitted signal is obtained. Then, the embedded binary number $I_n$ can be acquired according to the mapping relationship. It is worth noting that the sampling judgement moment is obtained from the detection process, so the information demodulation is conducted after the radar signal processing.

After the internal matched filtering, $K$ outputs are added together before the external matched filtering. According to (3), the sum is

$$r_{f1}(t) = \sum_{k=1}^{K} r_k(t) \\ = \sum_{n=0}^{N-1}\sum_{k=1}^{K} c_{n,k} R_{kk}(t-t_0-nT_s)e^{j\varphi_n} + w_{f1}(t) \quad (10)$$

where $w_{f1}(t) = \sum_{k=1}^{K} w_k(t)$. According to (2), the ideal autocorrelations of different orthogonal signals are the same, so $R_{kk}(t)$, $k=1,…,K$ in (10) can be replaced by $R(t)$, and (10) can be simplified to

$$r_{f1}(t) = R(t-t_0) \otimes h_{p2}(t) + w_{f1}(t) \quad (11)$$

where

$$h_{p2}(t) = \sum_{n=0}^{N-1} \delta(t-nT_s)e^{j\varphi_n} \quad (12)$$

in which $\delta(\cdot)$ is the Kronecker delta function. To simplify the following derivation, the steps mentioned above is rewritten as

$$r_{f1}(t) = r(t) \otimes h_{f1}(t) \quad (13)$$

where

$$h_{f1}(t) = \sum_{k=1}^{K} s_k^*(-t). \quad (14)$$

Subsequently, the external matched filtering is performed, and the output is

$$r_{f2}(t) = r_{f1}(t) \otimes h_{f2}(t) \quad (15)$$

where

$$h_{f2}(t) = h_{p2}^*(-t). \quad (16)$$

(15) can be transformed into

$$r_{f2}(t) = R(t-t_0) \otimes R_2(t) + w_{f2}(t) \quad (17)$$

where

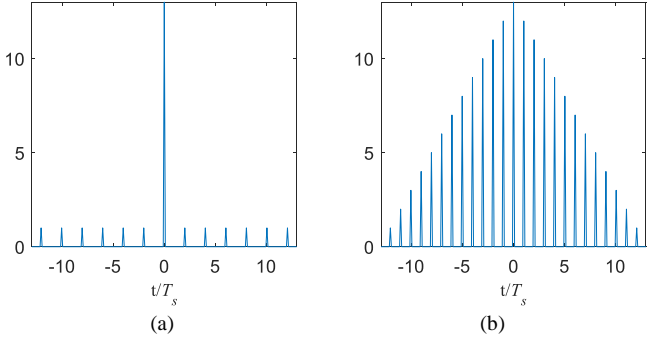

Fig. 4. (a) $R_2(t)$ of 13-Barker sequence. The PSL is 1/13. (b) $R_2(t)$ of $\mathbf{1}_{13\times 1}$ sequence. The PSL is 12/13.

$$w_{f2}(t) = w_{f1}(t) \otimes h_{f2}(t) \quad (18)$$

and

$$R_2(t) = h_{p2}(t) \otimes h_{f2}(t)$$
$$= \sum_{n=0}^{N-1}\sum_{m=0}^{N-1} \delta[t-(n-m)T_s]e^{j(\varphi_n - \varphi_m)}. \quad (19)$$

As shown in (19), the EP sequence determines the PSL of $R_2(t)$, so it is important to choose an appropriate EP sequence. Figure 4 shows if 13-Barker sequence is used as the EP sequence, the PSL of $R_2(t)$ is −22.3 dB. By contrast, if $\mathbf{1}_{13\times 1}$ is used as the EP sequence, the PSL is −0.7 dB.

The radar signal processing can be concluded as

$$r_{f2}(t) = r(t) \otimes h_f(t) \quad (20)$$

where

$$h_f(t) = h_{f1}(t) \otimes h_{f2}(t). \quad (21)$$

The signal processing procedure and the output of each step are shown in Figure 5. It can be found that the prior knowledge of the embedded information is not required for the radar signal processing mentioned above and $r_{f2}(t)$ is unrelated to the embedded information, which satisfy the requirement of the RUI. To evaluate the RUI performance of the proposed waveform, we define the dissimilarity (D) as

$$D = \max_{\forall t} \left| \frac{|y(t,\mathbf{I}_1)| - |y(t,\mathbf{I}_2)|}{E} \right| \quad (22)$$

where

$$y(t,\mathbf{I}) = s(t,\mathbf{I}) * h(t) \quad (23)$$

and $E$ is the energy of $s(t,\mathbf{I})$. $s(t,\mathbf{I})$ describes the JRC waveform embedded with information sequence $\mathbf{I}=[I_0\ I_1\ldots I_{N-1}]^T$ where $[\cdot]^T$ denotes the regular transpose operation and $h(t)$ is the corresponding radar signal processing matching function. It should be noted that the absolute value is used for the subtraction in (22), because amplitude detector is usually used in the detection process, and the detection performance depends on the amplitude of the output. Obviously, for the proposed waveform, the $D$ should be zero if signals are ideal orthogonal because $r_{f2}(t)$ remains the same with different $\mathbf{I}$. If signals are nonideal orthogonal, $r_{f2}$ will slightly change with different $\mathbf{I}$ and the $D$ will not be zero. The number of probable information sequences is $2^{NL}$, so the number of the probable $D$ values is $2^{2NL-1}+2^{NL-1}$. However, there is no need to traverse all the probabilities. With simulation times much smaller than $2^{2NL-1}+2^{NL-1}$, a stable mean of the $D$ can already be approached and be used to evaluate the RUI performance.

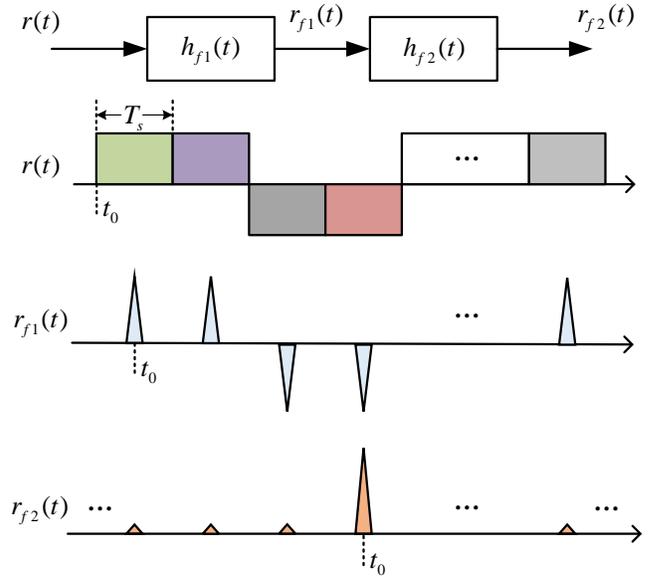

Fig. 5. Radar signal processing procedure and the corresponding output of each step.

## III. WAVEFORM PERFORMANCE ANALYSIS

### A. Detection performance

The peak signal to noise ratio (PSNR) of the output of the radar signal processing determines the detection performance. In the situation of white noise, the matched filtering method makes the PSNR the maximum [21]. However, the radar signal processing method in this paper is different from the matched filtering method, and it is necessary to analyze the PSNR of the output of the proposed method.

The output of the external matched filtering is used for radar detection, and its noise power is

$$\overline{w_{f2}^2(t)} = \frac{1}{2\pi}\frac{N_0}{2}\int_{-\infty}^{\infty}\left|H_f(\omega)\right|^2 d\omega$$
$$= \frac{N}{2\pi}\frac{N_0}{2}\int_{-\infty}^{\infty}\left|\sum_{k=1}^{K}S_k^*(\omega)\right|^2 d\omega \quad (24)$$
$$= \frac{NN_0}{2}\sum_{k=1}^{K}\sum_{j=1}^{K}R_{kj}(0)$$

where

$$H_f(\omega) = \sum_{k=1}^{K}S_k^*(\omega)\sum_{n=0}^{N-1}e^{j(\omega nT_s - \varphi_n)} \quad (25)$$

is the Fourier transform of $h_f(t)$. Then the PSNR of the output can be written as

$$d_2 = \frac{\int_{-\infty}^{\infty} s(t)h_f(-t)dt}{\overline{w_{f2}^2(t)}} = \frac{\left[\sum_{n=0}^{N-1}\sum_{k=1}^{K}\sum_{j=1}^{K}c_{n,k}R_{kj}(0)\right]^2}{\frac{NN_0}{2}\sum_{k=1}^{K}\sum_{j=1}^{K}R_{kj}(0)}. \quad (26)$$

Assuming the probability of each signal being transmitted is equal in one information symbol period $T_s$, the number of each signal being transmitted tends to be equal with the increase of the number of information symbol. Therefore, $c_{n,k}$ can be replaced by the probability of each signal being transmitted which is $1/K$ for the long term performance evaluation. Consequently, (26) can be simplified to

$$d_2 = \frac{\left[\frac{N}{K}\sum_{k=1}^{K}\sum_{j=1}^{K}R_{kj}(0)\right]^2}{\frac{NN_0}{2}\left[\sum_{k=1}^{K}\sum_{j=1}^{K}R_{kj}(0)\right]} = \frac{2N\sum_{k=1}^{K}\sum_{j=1}^{K}R_{kj}(0)}{K^2 N_0}. \quad (27)$$

If the signals are ideal orthogonal, (27) becomes

$$\overline{d_2} = \frac{2NKE_s}{K^2 N_0} = \frac{2E}{KN_0}. \quad (28)$$

In general, considering the nonideal orthogonality of the signals, (27) can be rewritten as

$$d_2 = \overline{d_2}(1+\gamma) \quad (29)$$

where

$$\gamma = \frac{1}{KE_s}\sum_{k=1}^{K}\sum_{\substack{j=1 \\ j\neq k}}^{K}R_{kj}(0). \quad (30)$$

$\overline{d_2}$ is $1/K$ of the PSNR of the output of the matched filtering method which is $2E/N_0$ with the signal energy of $E$ and the noise power spectral density of $N_0/2$. In conclusion, the PSNR of the output of the proposed method degrades with the increase of the number of bits embedded in an information symbol. Besides, $\gamma$ influences the PSNR of the output in practical systems, which leads to unstable detection performance. Therefore, it is necessary to control $\gamma$.

According to (30), $\gamma$ is decided by the sum of all the cross-correlation values at zero delay. From the perspective of detection, a negative $\gamma$ reduces the PSNR. To make $\gamma$ approach zero, the weighted cyclic algorithm-new in [22] can be used with more optimization weight put on the cross-correlation values at zero delay. Unfortunately, with extra optimization at zero delay, the optimization effect at other time delays might be sacrificed, leading to worse isolation (I) and higher PSL.

Thus, a new optimization method is proposed to solve this conflictation. With $K$ orthogonal signals $x_1$ to $x_K$ generated by methods which need no extra optimization weight on the cross-correlation values at zero delay, the optimized signal is represented as

$$s_k = x_k e^{j\alpha_k}, k = 1,...,K. \quad (31)$$

The objective function is

$$F = \gamma, \quad (32)$$

and the optimization variable is

$$\boldsymbol{\alpha} = [\alpha_1\ \alpha_2\ ...\ \alpha_K]^T. \quad (33)$$

The optimization problem can be described as

$$\max_{\boldsymbol{\alpha}\in[0,2\pi]^{K\times 1}} F(\boldsymbol{\alpha}). \quad (34)$$

It can be found that

$$\begin{aligned}\left|R_{kj}(t)\right| &= \left|\int_{-\infty}^{\infty} s_k(\tau)s_j^*(\tau-t)d\tau\right| \\ &= \left|e^{j(\alpha_k-\alpha_j)}\int_{-\infty}^{\infty} x_k(\tau)x_j^*(\tau-t)d\tau\right| \\ &= \left|\int_{-\infty}^{\infty} x_k(\tau)x_j^*(\tau-t)d\tau\right|,\end{aligned} \quad (35)$$

so the I and the PSL of the orthogonal signals remain steady during the optimization.

Genetic algorithm (GA) in [23] can be used to get a local optimal solution of (34). The principle of this optimization method is to utilize the energy of the cross-correlation at zero delay, making it beneficial to the accumulation of the peak value of the output. However, if the signals are ideal orthogonal, the energy of the cross-correlation at zero delay will be zero, and there is no space for the optimization. Thus, the optimization effect of the nonideal orthogonal signals exceeds that of the ideal orthogonal signals.

### B. Communication anti-noise performance

SER evaluates the anti-noise performance of the communication system, and it determines the application of the proposed waveform in a JRC system. Therefore, it is important to analyze the SER of the proposed system.

Figure 6 shows the demodulation procedure. After the detection, the range time delay $t_0$ is obtained, and the peak position sequence of $r_k(t)$, $k=1,...,K$ is

$$\mathbf{t} = [t_0\ t_0+T_s\ t_0+2T_s\ ...\ t_0+(N-1)T_s]^T \quad (36)$$

Sample $r_k(t)$ with $\mathbf{t}$, and the peak sequence is

$$\mathbf{r}_k = \sum_{j=1}^{K} R_{jk}(0)\cdot[c_{0,j}\ c_{1,j}\ ...\ c_{N-1,j}]^T \odot e^{j\boldsymbol{\varphi}} + \mathbf{w}_k \quad (37)$$

where $\boldsymbol{\varphi}=[\varphi_0\ \varphi_1...\varphi_n...\varphi_{N-1}]^T$ is known to the receiver and can be compensated, $\mathbf{w}_k$ is a $1\times N$ vector representing the noise sequence of the $k$-th channel and $\odot$ represents the Hadamard product. Subsequently, the comparison and judgement are made among the $K$ sequences from $\mathbf{r}_1$ to $\mathbf{r}_K$. The procedure mentioned above is defined as the coherent demodulation. The noise power of $\mathbf{w}_k$ is

$$\overline{w_k^2(t)} = \frac{1}{2\pi}\frac{N_0}{2}\int_{-\infty}^{\infty}\left|S_k^*(\omega)\right|^2 d\omega = \frac{N_0}{2}R_{kk}(0) = \frac{N_0 E_s}{2} \quad (38)$$

where $S_k(\omega)$ is the Fourier transform of $s_k(t)$ and $N_0/2$ is the noise power spectral density. Then the signal to noise ratio (SNR) of the information demodulation can be described as

$$d_1 = \frac{\left|R_{jj}(0)-R_{jk}(0)\right|^2}{N_0 E_s/2} \quad (39)$$

where $j$ is the serial number of the transmitted signal and $k$ is the serial number of the compared channel. If the transmitted signals are well orthogonal, $R_{jk}(0)$ can be ignored when $k\neq j$ and (39) can be simplified to

$$d_1 \approx \frac{\left|R(0)\right|^2}{N_0 E_s/2} = \frac{2E_s}{N_0}. \quad (40)$$

The SER of the coherent demodulation is [24]

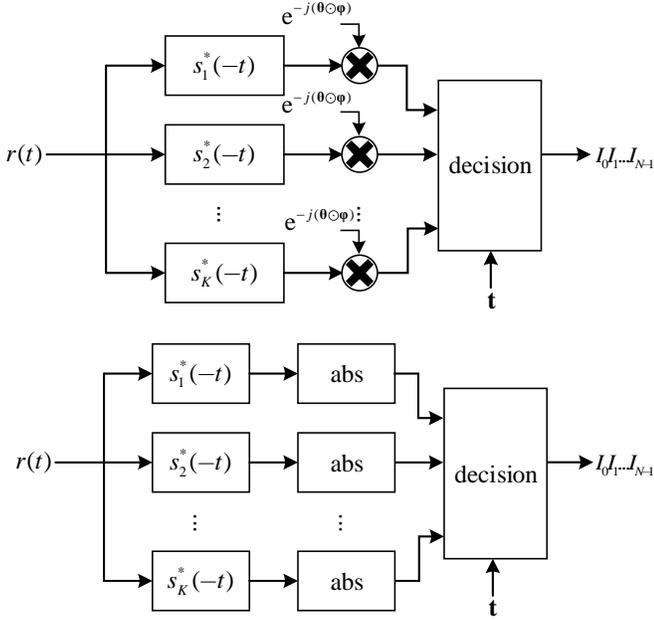

Fig. 6. Coherent demodulation (above) and non-coherent demodulation (below).

$$P_e = 1 - \int_{-\infty}^{\infty} \frac{e^{-u^2/2}}{\sqrt{2\pi}} \left[ \int_{-\infty}^{u+\sqrt{2d_1}} \frac{e^{-z^2/2}}{\sqrt{2\pi}} dz \right]^{K-1} du \quad (41)$$

which is also the SER of the coherent demodulation of the MFSK system.

Nevertheless, an uncertain phase shift $\theta$ occurs at the propagation of the emission, so the peak sequence is rewritten as

$$\mathbf{r}_k = \sum_{j=1}^{K} R_{kj}(0) \cdot \left[ c_{0,k}\ c_{1,k}\ ...\ c_{N-1,k} \right]^{\mathrm{T}} \odot e^{j\varphi} \odot e^{j\theta} + \mathbf{w}_k. \quad (42)$$

Note that $\theta$ can be compensated by the knowledge acquired from the detection. However, this process raises the complexity of the demodulation system. Instead, the comparison and the judgement of the non-coherent demodulation are made on the amplitude of $\mathbf{r}_k$, which eliminate the influence of $\theta$. The SER of the non-coherent demodulation is [25]

$$P_e \le \frac{K-1}{2} e^{-d_1/4} \quad (43)$$

which is also the SER of the non-coherent demodulation of the MFSK system. Compared with the coherent demodulation, the non-coherent demodulation raises the SNR requirement slightly, while it reduces the complexity of the demodulation system.

### C. Doppler performance

Ambiguity function (AF) is a common tool to design and analyze radar waveform [21]. For signal $u(t)$ and its echo with time delay $t$ and Doppler shift $f_d$, the AF of $u(t)$ is represented as

$$\overline{A}(t, f_d) = \int_{-\infty}^{\infty} u(x) \exp(j2\pi f_d x) u^*(x-t) dx. \quad (44)$$

However, for the proposed radar signal processing method, $u^*(t)$ is unknown to the receiver because of the information embedded in the waveform. Thus the AF of the proposed waveform is defined as

$$A(t, f_d) = \int_{-\infty}^{\infty} s(x) \exp(j2\pi f_d x) h_f(t-x) dx. \quad (45)$$

Assuming $T_s f_d \ll 1$, as derived in Appendix, (45) can be simplified to

$$A(t, f_d) = R(t, f_d) \otimes R_2(t, f_d) \quad (46)$$

where $R(t, f_d)$ is the AF of the ideal orthogonal signal and $R_2(t, f_d)$ is the AF of $h_{p2}(t)$. With a certain Doppler frequency $f_0$ which satisfies $T_s f_0 \ll 1$, $R(t, f_0)$ resembles an impulse signal compared with $R_2(t, f_0)$, which indicates that the PSL of $A(t, f_0)$ mainly depends on the PSL of $R_2(t, f_0)$. Generally, the Doppler performance of the proposed waveform mainly depends on the EP sequence.

If 13-Barker sequence is used as the EP sequence, the range sidelobe will rise significantly with the increase of the Doppler frequency, which may cause false alarm in the process of detection. This problem can be solved by optimizing the Doppler performance of the EP sequence. The objective function can be described as

$$F = \max_{\substack{f_d \in [f_1, f_2], f_d \ne 0 \\ t \in [t_1, t_2], t \ne 0}} \left| R_2(t, f_d) \right| \quad (47)$$

where $t_1$ to $t_2$ is the time range to be optimized and $f_1$ to $f_2$ is the Doppler frequency range to be optimized. The optimization problem is described as

$$\min_{\varphi \in [0, 2\pi]^{N \times 1}} F(\varphi). \quad (48)$$

A number of waveform design methods can be used for the optimization mentioned above [26]-[28]. In the next section, the Doppler performance of the proposed waveform with different EP sequences will be compared.

## IV. SIMULATION RESULTS

In our simulation, the cyclic algorithm-new proposed in [22] is used to generate the $K$ orthogonal signals $x_1$ to $x_K$ with length $M$, and $x_1$ to $x_K$ are optimized in the way of (34) to generate $s_1$ to $s_K$ which are the orthogonal signals applied in the internal modulation.

### A. Detection probability

To show the detection performance of the proposed radar signal processing method, we analyze the detection probability ($P_d$) of the proposed method by both Monte-Carlo simulations and theoretical calculation and compare it to that of the matched filtering method. Besides, the relationship between the detection performance and the communication performance is shown by analyzing the $P_d$ with different numbers of bits embedded in a pulse.

The comparison between the performance of the proposed method and the matched filtering method is shown in Figure 7 where the X-axis is $d=2E/N_0$. The number of the orthogonal signals is $K=2$, $K=4$ and $K=8$, respectively, the length of the orthogonal signals is $M=200$, the constant false alarm probability is $10^{-6}$ and the times of the Monte-Carlo simulations are $10^8$. 13-Barker sequence is used as the EP sequence. The performance of the proposed method with the optimized nonideal orthogonal signals is simulated and the performance of the proposed method with the ideal orthogonal signals is calculated. It can be found that under a certain level of $P_d$, the required $d$ of the proposed method with ideal orthogonal signals is $K$ times larger than that of the matched filtering method. In contrast, the required $d$ of the proposed method with the optimized nonideal orthogonal signal decreases by 0.1, 1 and 1.7 dB compared with that of the proposed method with ideal orthogonal signals when $K=2$, $K=4$ and $K=8$, respectively. Apparently, the optimization effect is improved with the increase of $K$, which is because the number of the optimization variables is the same as $K$, and the increase of $K$ provides more freedom for the optimization. Besides, the orthogonality degrades with the growth of $K$, and the cross-correlation values at zero delay tend to increase, which provides larger space for the optimization. Generally, the SNR requirement of the proposed method is larger than that of the matched filtering method to achieve a same level of $P_d$. Fortunately, the proposed optimization method can reduce the SNR requirement and the optimization effect is improved with the increase of $K$.

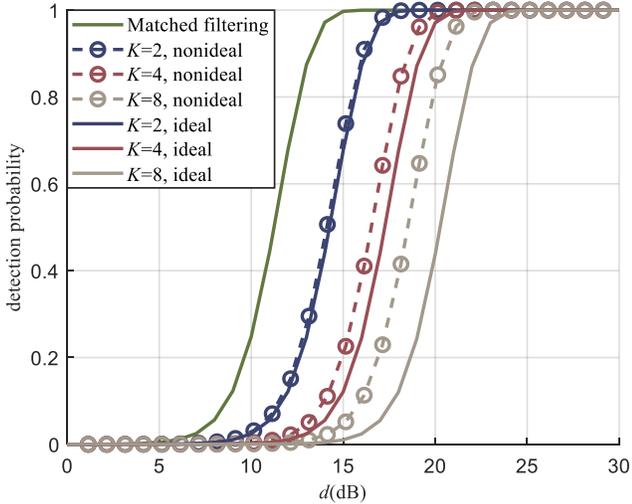

Fig. 7. Detection probability of the proposed method with ideal orthogonal signals and nonideal orthogonal signals with different $K$, compared with that of the matched filtering method.

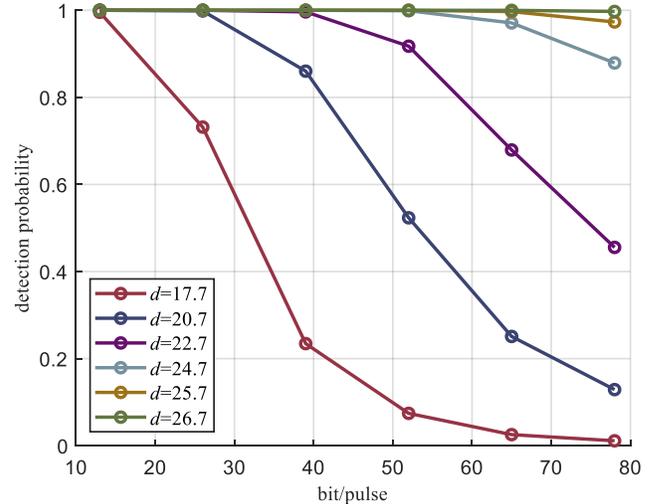

Fig. 8. Detection probability as a function of number of bits in a pulse for various SNRs.

The $P_d$ with different numbers of bits embedded into a pulse is shown in Figure 8. The number of bits which is controlled by $K$ is 13, 26, 39, 52, 65 and 78, respectively, the length of the orthogonal signals is $M=400$, the constant false alarm probability is $10^{-6}$ and the times of the Monte-Carlo simulations are $10^7$. 13-Barker sequence is used as the EP sequence. It can be found in Figure 8 that the $P_d$ experiences a decrease with the increase of the number of bits, and $d$ needs to be improved to maintain a high level of $P_d$. To reach a $P_d$ higher than 99%, the required $d$ is 17.7, 20.7, 22.7, 24.7, 25.7 and 26.7 dB with 13, 26, 39, 52, 65 and 78 bits embedded in a pulse, respectively. Apparently, the increase of the required $d$ decelerates with the growth of the number of bits, which is because the optimization effect is improved as $K$ ascends. In conclusion, the detection performance is restricted by the amount of information embedded in a pulse with constant power, bandwidth and duration, and the adjustment of the balance is realized by using different $K$. Moreover, the proposed optimization method is able to weaken the restriction between the detection performance and the amount of information embedded in a pulse.

### B. Symbol error rate

We simulate the SER of the coherent demodulation and the non-coherent demodulation by Monte-Carlo simulations to analyze the communication performance of the proposed waveform. The number of the orthogonal signals is $K=2$, $K=4$ and $K=8$, respectively, the length of the orthogonal signals is $M=200$ and the times of the Monte-Carlo simulations are $10^7$. The simulation results are compared with the theoretical value which is also the SER of the MFSK system.

The performance of the coherent modulation is shown in Figure 9. Under a certain level of SER, the signal to noise ratio per bit ($r_b$) which equals to $2E_s/N_0/L$ decreases if $K$ ascends, which indicates that the efficiency of information transmission is improved with the increase of the number of bits embedded in a symbol. The simulation result is close to the theoretical result. However, nuances between them can be noticed. Under a certain level of SER, the required $r_b$ of the simulation is slightly higher than that of the theoretical result, which is because the cross-correlation is ignored for the simplification in (40), and the actual SNR is slightly lower than $2E_s/N_0$. Moreover, the cross-correlation values at zero delay tend to ascend because of the optimization in (34), and the difference could be enhanced. Fortunately, the value is still small compared with the peak value of the autocorrelation of the orthogonal signals. As a result, the SNR requirement modestly increases no more than 0.5 dB than that of the coherent demodulation of the MFSK system for a SER no less than $10^{-5}$.

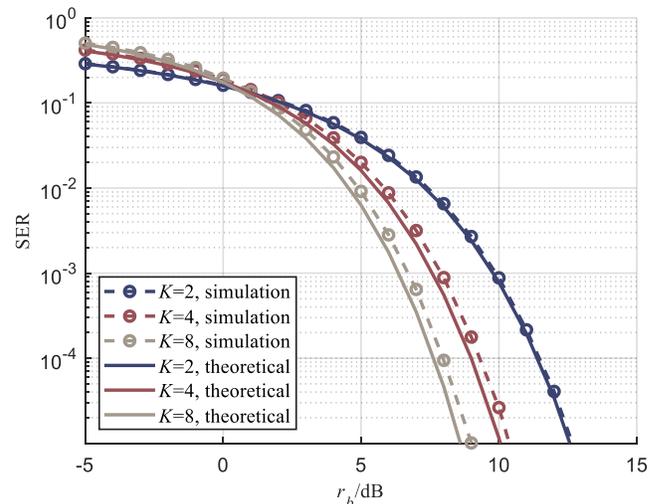

Fig. 9. SER of the coherent demodulation as a function of signal to noise ratio per bit.

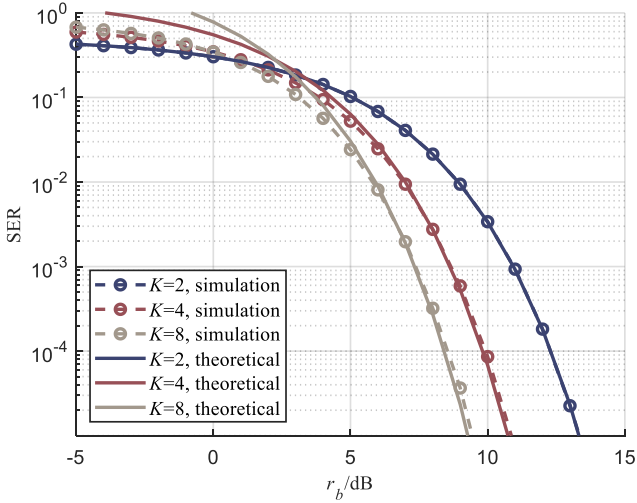

Fig. 10. SER of the non-coherent demodulation as a function of signal to noise ratio per bit.

Figure 10 shows the SER of the non-coherent demodulation. Note that the upper bound of the theoretical SER is drew in Figure 10 instead of the accurate value because the latter is hard to calculate, and the difference is negligible in the situation of high SNR [25]. To achieve a SER of $10^{-5}$, the required $r_b$ is 0.5, 0.3 and 0.4 dB higher than that of the coherent demodulation when $K=2$, $K=4$ and $K=8$, respectively. The difference between the simulation result and the theoretical result is negligible compared with that of the coherent demodulation because the comparison and the judgement of the non-coherent demodulation is based on the amplitude value which does not change with the optimization in (34). In conclusion, the SER performance of the non-coherent demodulation of the proposed method matches with that of the non-coherent demodulation of the MFSK system, and the SNR requirement increases no more than 0.5 dB than that of the coherent modulation of the proposed method for a SER no less than $10^{-5}$.

### C. Ambiguity function

The Doppler performance of the proposed waveform is displayed by the simulation of the AF. The number of the orthogonal signals is $K=2$ and the length of the orthogonal signals is $M=200$. Note that $M$ should be set according to the demand of the orthogonality to make sure the autocorrelation of the orthogonal signals can be considered similar to an impulse and the simplification in (46) is valid.

The AF of the orthogonal signal, the EP sequence and the proposed waveform is simulated, respectively. Shown in Figure 11 (a), there is a limited Doppler effect on the correlation of the orthogonal signal, and the autocorrelation of the orthogonal signal resembles an impulse signal when $|Tf_d|\leq 0.2$. Comparing Figure 11 (c) with Figure 11 (b), there is a remarkable resemblance between the two figures, which proves the conclusion in (46).

To further demonstrate the conclusion in (46), the PSLs of $A(t, f_d)$ and $R_2(t, f_d)$ are compared with different $f_d$. Meanwhile, the GA-optimized sequence and 13-Barker sequence is used as the EP sequence, respectively. The optimization time delay range is $-\infty \leq t \leq +\infty$ and Doppler frequency range is $-f_d \leq f \leq f_d$. Demonstrated in Figure 12, the PSL of $R_2(t, f_d)$ matches with that of $A(t, f_d)$ except the point of 13-Barker sequence at $Tf_d=0$. At this point, the PSL of $R(t, f_d)$ is comparable to that of $R_2(t, f_d)$, which leads to the increase of the PSL of $A(t, f_d)$. With the growth of $f_d$, the PSL of $R_2(t, f_d)$ rises and far exceeds that of $R(t, f_d)$ which remains stable, and the PSL of $A(t, f_d)$ is

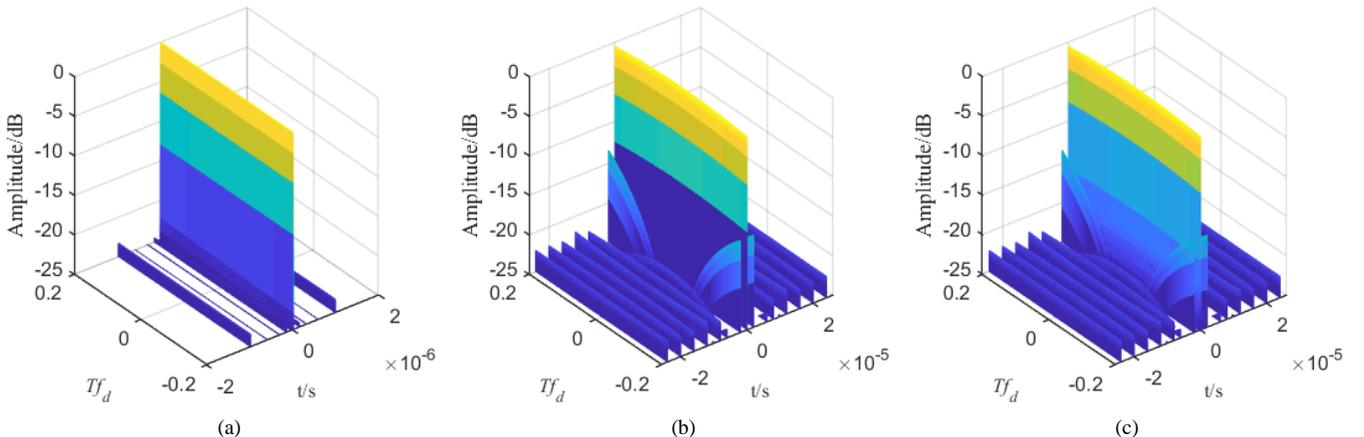

(a) (b) (c)
Fig. 11. (a) The AF of the orthogonal signal $R(t, f_d)$. (b) The AF of the EP sequence $R_2(t, f_d)$. (c) The AF of the proposed waveform $A(t, f_d)$.

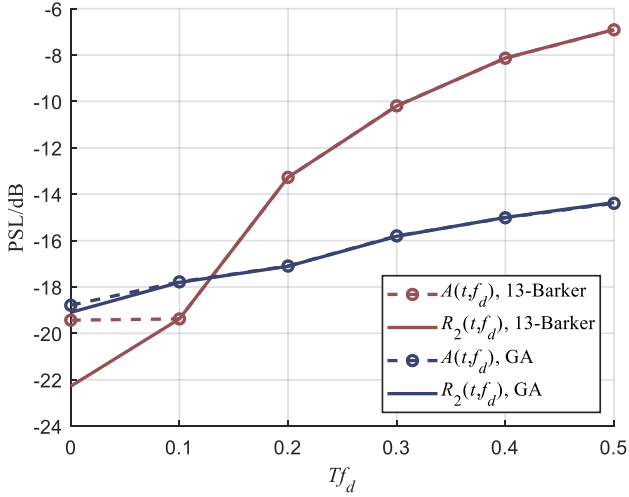

Fig. 12. PSL comparison with different max Doppler frequency.

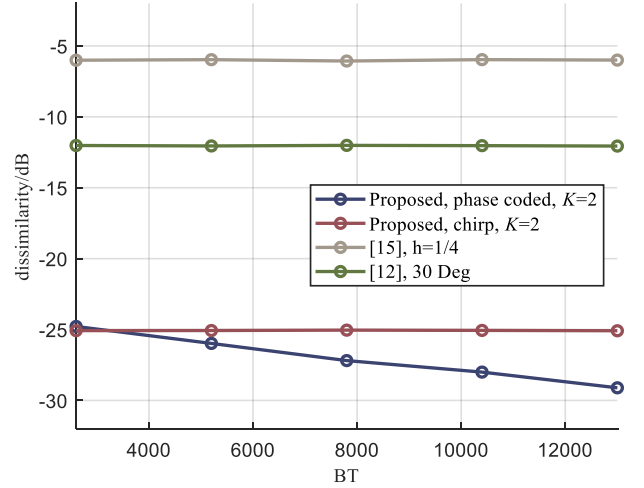

Fig. 13. RUI performance for different methods described by dissimilarity.

determined by the PSL of $R_2(t, f_d)$. Comparing the performance of the proposed waveform with different EP sequences, the PSL of $A(t, f_d)$ with 13-Barker sequence is lower than that with the GA-optimized sequence when $Tf_d \leq 0.1$. However, it goes up dramatically from −19 to −7 dB with the increase of $Tf_d$. In contrast, the PSL of $A(t, f_d)$ with the GA-optimized sequence, although witnessing a modest increase, is always lower than −14 dB. In conclusion, The Doppler performance of the proposed waveform is decided by the Doppler performance of the EP sequence, and it can be improved by the optimization of the Doppler performance of the EP sequence.

*D. Dissimilarity*

In order to show the RUI performance of the proposed waveform, we calculate the $D_{\text{mean}}$ of the proposed waveform by Monte-Carlo simulations. The $D_{\text{mean}}$ is described as

$$D_{\text{mean}} = 20\log \frac{\sum_{m=1}^{Mon} \max_{\forall t} \frac{\left| |y(t, \mathbf{I}_{m1})| - |y(t, \mathbf{I}_{m2})| \right|}{E}}{Mon} \quad (49)$$

where $\mathbf{I}_{m1}$ and $\mathbf{I}_{m2}$ are generated randomly and the times of the Monte-Carlo simulations are $Mon$=5000. 13-Barker sequence is used as the EP sequence.

The waveforms in [12] and [15] are analyzed for the comparison. The matched filtering method is used for the radar signal processing in these two methods, so the matching function of them can be described as

$$h(t) = s^*(-t, \mathbf{I}_{m1}). \quad (50)$$

For the sake of fairness, all three waveforms are embedded with 13 bit information and share the same time-bandwidth products (BT), and the SER performances are controlled at the same level by making the phase change $\phi_\delta$=30° in [12], the modulation index h=1/4 in [15] and the number of the orthogonal signals $K$=2 in our method. Besides, we simulate the performance of the proposed waveform with different groups of orthogonal signals, including the up chirp and the down chirp signals and the phase-coded signals.

As shown in Figure 13, the $D_{\text{mean}}$ of [12], [15] and the proposed waveform based on the up chirp and the down chirp signals is around −6, −12 and −25 dB, respectively. All three of them remain stable with the increase of BT. It is apparent that the $D_{\text{mean}}$ of the proposed waveform is much smaller than those of other waveforms, indicating that the RUI performance of the proposed waveform is better than those of others. As we analyzed in Section II-C, $D$ of the proposed waveform will descend with the increase of the orthogonality of the orthogonal signals. For the up chirp and the down chirp signals, the I decreases but the PSL remains the same with the growth of BT, so the increase of the orthogonality is limited. In contrast, the $D_{\text{mean}}$ of the proposed method based on the phase coded signals drops with the growth of BT. It starts at −24 dB and decreases to −29 dB with a BT increasing from 2600 to 13000. The $D_{\text{mean}}$ decreases because both the PSL and the I of the phase coded signals decrease with the growth of BT (code length), which leads to the increase of the orthogonality. It can be concluded that the RUI performance of the proposed waveform with well orthogonal signals is better than those of the waveforms in [12] and [15], and the performance is improved with the ascendance of the orthogonality of the orthogonal signals applied in the internal modulation.

V. CONCLUSION

In this paper, we have presented a unimodular JRC waveform realized by composite modulation. The embedded information is not required for the corresponding signal processing, and the detection performance is robust to unknown information. The communication performance and the detection performance depend on the number of the orthogonal signals applied in the internal modulation. Specially, the bit rate is $N\log_2 K$ per pulse with a SER similar to that of the MFSK system and the PSNR of the output of the radar signal processing is limited by $K$. An optimization method is proposed to weaken the restriction by utilizing the nonideal orthogonality of the signals. The Doppler performance of the proposed waveform depends on the EP sequence applied for the external modulation. Future work can extend the waveform to the multi-target scenario, especially addressing the design of multi-target detection and channel equalization.

## APPENDIX

The derivation of the AF of the proposed waveform is as follows.

The AF is defined as

$$A(t, f_d) = \int_{-\infty}^{\infty} s(x) e^{j2\pi f_d x} h_f(t-x) dx$$

$$= \int_{-\infty}^{\infty} \begin{bmatrix} e^{j2\pi f_d x} \sum_{n=0}^{N-1} \sum_{k=1}^{K} c_{n,k} s_k(x-nT_s) e^{j\varphi_n} \\ \cdot \sum_{n=0}^{N-1} \sum_{k=1}^{K} s_k^*(x-t-nT_s) e^{-j\varphi_n} \end{bmatrix} dx$$

$$= \int_{-\infty}^{\infty} \left\{ \sum_{n=0}^{N-1} \sum_{m=0}^{N-1} \sum_{k=1}^{K} \sum_{j=1}^{K} \begin{bmatrix} e^{j2\pi f_d x} c_{n,k} s_k(x-nT_s) \\ \cdot s_j^*(x-t-mT_s) e^{j(\varphi_n-\varphi_m)} \end{bmatrix} \right\} dx.$$
(51)

Let $x_0 = x - nT_s$, and (51) can be transformed into

$$A(t, f_d) = \int_{-\infty}^{\infty} \left\{ \sum_{n=0}^{N-1} \sum_{m=0}^{N-1} \sum_{k=1}^{K} \sum_{j=1}^{K} \begin{bmatrix} e^{j2\pi f_d x_0} c_{n,k} s_k(x_0) \\ \cdot s_j^*[x_0-t-(m-n)T_s] \\ \cdot e^{j2\pi f_d nT_s} e^{j(\varphi_n-\varphi_m)} \end{bmatrix} \right\} dx_0$$

$$= \sum_{n=0}^{N-1} \sum_{m=0}^{N-1} \sum_{k=1}^{K} \sum_{j=1}^{K} \left\{ \begin{array}{l} c_{n,k} R_{kj}[t-(n-m)T_s, f_d] \\ \cdot e^{j2\pi f_d nT_s} e^{j(\varphi_n-\varphi_m)} \end{array} \right\}$$
(52)

where

$$R_{kj}(t, f_d) = \int_{-\infty}^{\infty} e^{j2\pi f_d x} s_k(x) s_j^*(x-t) dx.$$
(53)

According to [20], a basic requirement of limited Doppler effect on the correlations of the signals is

$$T_s f_d \ll 1.$$
(54)

If the requirement is satisfied, according to (2) and (3), (53) can be simplified to

$$\begin{array}{l} R_{kk}(t, f_d) = R(t, f_d), k = 1, ..., K \\ R_{kj}(t, f_d) = 0, k \neq j, j = 1, ..., K \end{array}$$
(55)

where $R(t, f_d)$ is the AF of the ideal orthogonal signal. Substituting (55) into (52), it can be obtained that

$$A(t, f_d) = \sum_{n=0}^{N-1} \sum_{m=0}^{N-1} \left\{ \begin{array}{l} R[t-(n-m)T_s, f_d] \\ \cdot e^{j2\pi f_d nT_s} e^{j(\varphi_n-\varphi_m)} \end{array} \right\}$$

$$= R(t, f_d) \otimes \sum_{n=0}^{N-1} \sum_{m=0}^{N-1} \left\{ \begin{array}{l} \delta[t-(n-m)T_s] \\ \cdot e^{j2\pi f_d nT_s} e^{j(\varphi_n-\varphi_m)} \end{array} \right\}$$

$$= R(t, f_d) \otimes \begin{bmatrix} \sum_{n=0}^{N-1} e^{j2\pi f_d nT_s} \delta(t-nT_s) e^{j\varphi_n} \\ \otimes \sum_{m=0}^{N-1} \delta(-t-mT_s) e^{-j\varphi_m} \end{bmatrix}$$

$$= R(t, f_d) \otimes \left[ e^{j2\pi f_d t} h_{p2}(t) \otimes h_{p2}^*(-t) \right]$$

$$= R(t, f_d) \otimes R_2(t, f_d).$$
(56)